\documentclass[aps,groupedaddress, 12pt]{revtex4-2}
\usepackage[dvipsnames]{xcolor}
 \usepackage{graphicx}
 \usepackage{amsmath} 
\usepackage{eucal} 
\usepackage{amssymb} 
\usepackage{cancel}
\usepackage[normalem]{ulem}
\usepackage{amsmath}
\newcommand{\stkout}[1]{\ifmmode\text{\sout{\ensuremath{#1}}}
\else\sout{#1}\fi}

\usepackage{mathptmx}

\begin{document}
\title{ A visual proof of entropy production 
during thermalization with a heat reservoir}
 \author{Ramandeep S. Johal} 
 \email[e-mail: ]{rsjohal@iisermohali.ac.in}
 \affiliation{ Department of Physical Sciences, 
 Indian Institute of Science Education and Research Mohali,  
 Sector 81, S.A.S. Nagar, Manauli PO 140306, Punjab, India} 
 \begin{abstract}
In this note, the equilibrium curve of a thermodynamic
system is used to depict entropy production in the process
of thermalization with a reservoir. For the given 
initial and final equilibrium states of the system, the entropy
production is reduced when work is also 
extracted during thermalization. 
The case of maximum work extraction corresponds to a reversible
process. For less than optimal work extraction, 
the lost available work is shown to be directly
proportional to the entropy produced. 
 \end{abstract}
\maketitle
\par\noindent 
 
Entropy production is the core concept underlying
the Second law which states that 
irreversible or spontaneous processes always
increase the entropy of the universe. 
An example is the flow of heat across a temperature
gradient such as when 
a system thermalizes with a heat reservoir. Here,
even if the heat is transferred in a 
quasi-static manner, there is a net increase
in the total entropy of the system {\it plus}
reservoir \cite{Callenbook}. 
For the case when  the reservoir 
is initially at a higher (lower) temperature than the system,
it implies that the increase (decrease) in the entropy
of the system is more than 
the decrease (increase) in the entropy of the reservoir.
Now, the change in the entropy
of the system depends on its nature, unlike  
for the reservoir. Assuming an ideal gas
system, the increase of total entropy
may be easily demonstrated using 
the well-known logarithm inequality. Visual demonstrations
of the Second law for such irreversible processes 
have also assumed an ideal-gas type behavior
for the system \cite{Bucher1993, Vallejo24a}. 
In this paper, we present a diagram 
using equilibrium curve of the 
system which overcomes these limitations.
Since visual proofs often 
help in easy comprehension of abstract
concepts, a demonstration of entropy
production based on a generic thermodynamic system 
is desirable.

For a given amount of a
thermodynamic 
system, the equilibrium state is described in terms of its internal energy $U(S,V)$ as a function 
 of its entropy $S$ and volume $V$  \cite{Callenbook}. 
Then, the temperature of the 
system is defined as $T = (\partial U/ \partial S)_{V}^{}$. One of the fundamental attributes 
of the equilibrium state is that $U(S,V)$ is a convex function of $S$ at 
constant $V$, which implies that the heat capacity
at constant volume is positive ($C_V > 0$).  
Now, with an initial state at energy $U_1$, entropy $S_1$ and 
temperature $T_1$, the system is placed in thermal contact
with a heat reservoir at temperature $T_2 > T_1$.
Heat flows from the reservoir to the system
till its temperature rises to $T_2$, corresponding
to a final energy $U_2$ and entropy $S_2$.
As the temperature is defined to be positive, 
$U_2 > U_1$ implies $S_2 > S_1$, for a fixed volume.

Thus, the system is in thermodynamic equilibrium 
in the 
 initial and the final state of the process. 
The entropy of the system increases by $\Delta S = S_2-S_1$,  while 
the energy increases by  $\Delta U = U_2 - U_1$ which
equals the heat exchanged with the reservoir, 
$Q_2= Q_1 = \Delta U$. On the other hand, 
the decrease in the entropy of the reservoir is:
$\Delta S_{\rm res} = Q_2/T_2$. 
Thus, the net or total change in the entropy of 
system {\it plus} reservoir is given by:
\begin{equation}
 \Delta S_{\rm tot} = \Delta S - \Delta S_{\rm res}.
 \label{dstot}
\end{equation}
The standard evaluation of the above quantities goes as follows.   
As the process happens  at a fixed 
system volume $V$, we can write
$\Delta U = \int_{T_1}^{T_2} C_V dT$ and  
$\Delta S = \int_{T_1}^{T_2} (C_V / T) dT$. 
Without assuming a specific form for
the function $C_V(T) >0$, 
 a general proof showing 
 $\Delta S_{\rm tot} >0$ is 
as follows. From the explicit expressions
given above, we can 
write Eq. (\ref{dstot}) as
\begin{equation}
 \Delta S_{\rm tot} = \int_{T_1}^{T_2} 
 \left( \frac{1}{T} - \frac{1}{T_2}  \right)
 C_V \; dT.
 \label{dstotin}
\end{equation}
Since $T_2 > T_1$, the integrand above is positive and 
so is the value of the integral. A similar proof can be 
constructed for the case $T_2 < T_1$ i.e. when 
the system is cooled by the reservoir.

\begin{figure}[ht]
 \includegraphics[width=9cm]{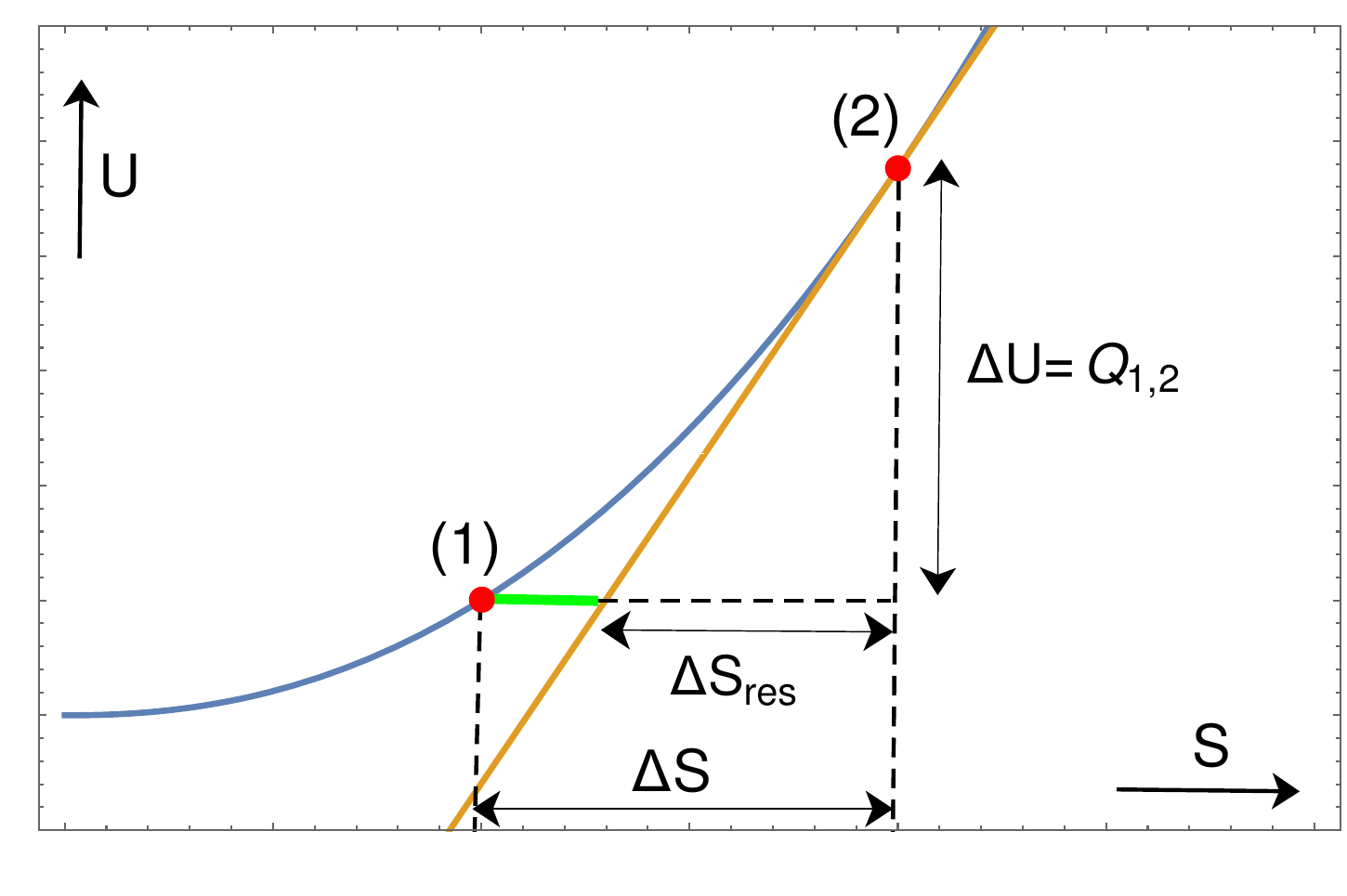}
 \caption{
 Points (1) and (2) (red online)  respectively denote the initial 
 and final states of the system in the 
 $\mbox{S-U}$ plane. The curved line (blue) 
 is the equilibrium curve which the system 
 may not follow during the thermalization process.
 The reservoir temperature
 $T_2$ is the slope of the tangent
 to the curve 
 at point 2. From the tangent and the $\Delta U$ segment,
the decrease in the entropy of the reservoir 
is depicted as $\Delta S_{\rm res} = \Delta U/T_2$.
 The increase in the entropy of the system is 
 given by $\Delta S = S_2-S_1$. Clearly, $\Delta S >
 \Delta S_{\rm res}$, and the green segment
 of length ($\Delta S - \Delta S_{\rm res}$)
 denotes entropy production in the process.
 }
 \label{fig1}
\end{figure}
Fig. 1  shows the (convex) equilibrium curve $U(S)$ 
of the system at a  given volume $V$. The entropy
changes involved in the process are 
depicted as certain line segments  
showing that $\Delta S > \Delta S_{\rm res}$, 
and hence $\Delta S_{\rm tot}>0$ due to Eq. (\ref{dstot}). 
Further, the diagram is only based on 
two properties: i) positivity of the temperature
and ii) convexity of the function $U(S)$.
A major difference of the present diagram from
the previous ones is that it is not restricted
to the ideal gas systems. 
The reader is invited 
to draw the corresponding diagram 
for the case where the reservoir is at a lower
temperature than the system ($T_2 < T_1$). 

Note that the net rise in total entropy does
not require a complete thermalization with
the reservoir. Any amount of heat flow 
across a finite temperature gradient 
increases the  total entropy. 
Following Fig. 1,
we can as well analyze the case of incomplete thermalization
where the final state of the system is some intermediate
state lying on the equilibrium curve in between 
the points (1) and (2). 
The temperature of the system $T'$ ($T_1 < T' < T_2$)
is again given by the slope of the tangent at that point.
It is easily seen that 
$\Delta S_{\rm tot} >0$ holds in this case too,
though the length of the segment denoting 
entropy production is smaller than in the case
of complete thermalization. Thus, 
we observe from the figure that the entropy
production attains its maximum value when the 
system reaches thermal equilibrium with the 
reservoir. 

\begin{figure}[ht]
 \includegraphics[width=9cm]{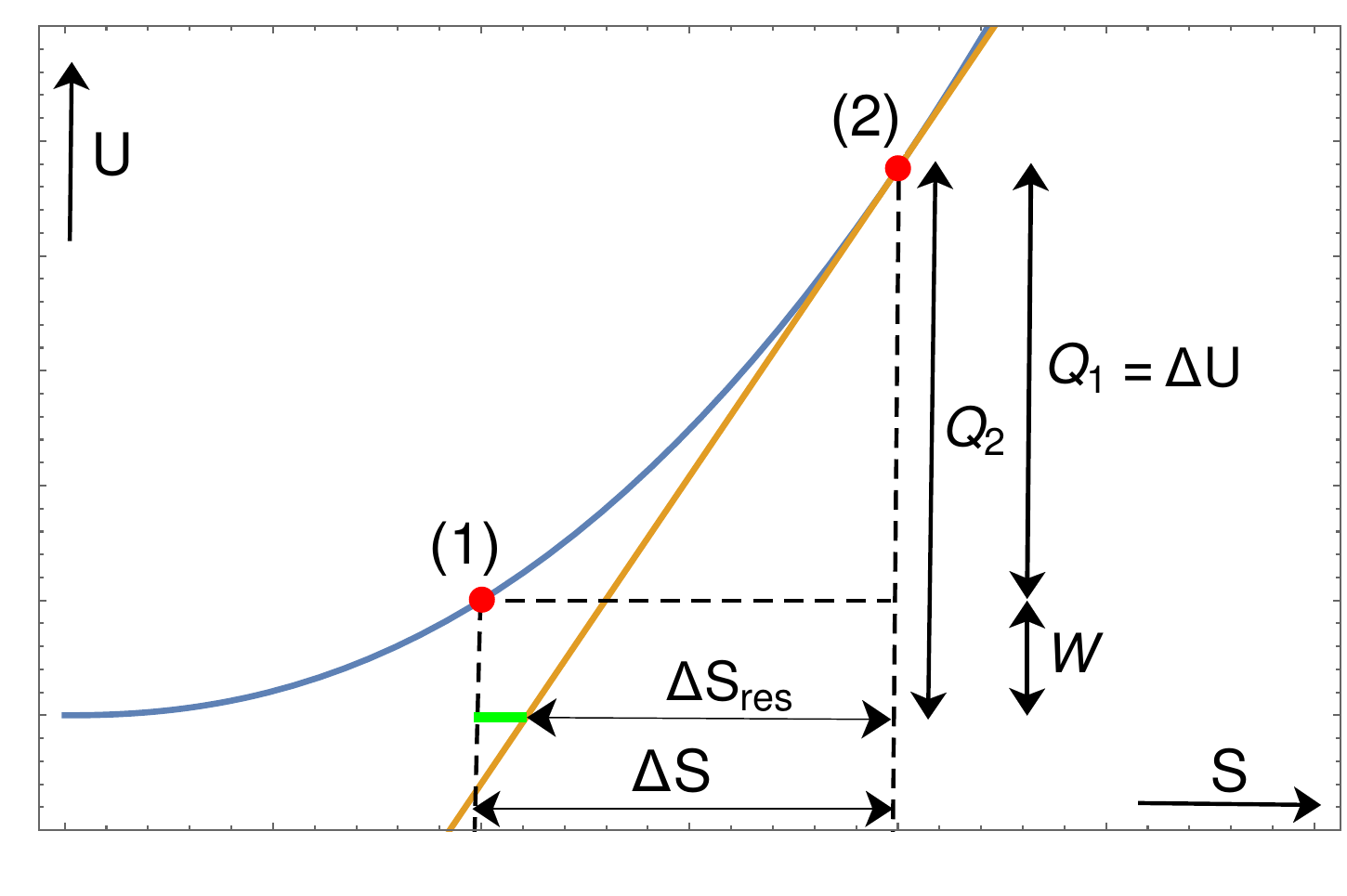}
 \caption{Thermalization with a reservoir
 in the presence of work extraction $W$. The 
 system absorbs the same amount of heat 
 as $Q_1$ in Fig. 1, but the heat absorbed
 from the reservoir is $Q_2 = Q_1 + W$. By comparing
 the lengths of the green segments in the two figures, we note that less amount of entropy is produced
 if some work is extracted. The work 
 extracted is maximum when the length of the 
 green segment shrinks to zero, corresponding to 
 a reversible process.}
 \label{fig2}
\end{figure} 
Suppose that instead of making a thermal contact, we couple the reservoir
and the system by means of a heat engine 
for which these 
act as heat source and heat sink, respectively.
The engine runs by executing certain heat 
cycles---absorbing an amount of heat from the reservoir, converting
a part of it into work and rejecting the 
rest of the heat to the system. The engine 
produces useful work till the system comes to be
in thermal equilibrium with the reservoir. 
Thus, the initial and the final states
of the system are the same as in the case of 
thermalization above, yielding 
the amount of heat rejected to the sink as 
$Q_1 = \Delta U$. Likewise, the change in system entropy
is equal to  $\Delta S$. Now, suppose that $W \geq 0$ amount 
of work is extracted by the end of this process.
Since the engine undergoes cycles,
the conservation of energy implies that 
$Q_2 = W+Q_1$ amount of heat is absorbed from 
the reservoir, which implies  $Q_2 \geq Q_1$.
This process is depicted in Fig. 2. 
It is apparent that the entropy production here
is smaller in magnitude as compared to pure thermalization
where no work was extracted.  
Fig. 2 also suggests that the magnitude of
work can be enhanced till
$\Delta S_{\rm res} = \Delta S_{}$ i.e.  
when $\Delta S_{\rm tot}$ vanishes and 
the engine becomes a reversible one.
Thus, we observe that 
maximum work ($W_{\rm max}$)  
is extracted when 
thermalization proceeds as 
a reversible process---with no entropy
production. In general, we have
$W \leq W_{\rm max}$. In fact, using the similarity property of triangles in Fig. 2,
we can show that $W_{\rm max} - W = T_2\Delta S_{\rm tot}$,
where $\Delta S_{\rm tot}$ is the entropy produced in the 
process that extracts $W$ amount of work. The difference 
$W_{\rm max} - W$, called the lost available work or 
the exergy destroyed, is directly proportional to 
the entropy produced. This relation is 
well known in the engineering parlance as 
the Gouy-Stodola theorem. 
Since, the initial and 
final states of the system remain the same
irrespective of the amount of work extracted, 
so it implies that $Q_2 = W + \Delta U$ increases 
in direct proportion to the work extracted, 
with its maximum value being $T_2 \Delta S$
(see Fig. 2). 
Thus, we obtain $W_{\rm max}  = T_2 \Delta S - \Delta U$.

Temperature-energy interaction diagrams, depicting 
heat and work flows in reversible as well as irreversible heat cycles,  
were introduced in the engineering
literature \cite{Brodinskii1973, Bejan1977} and 
also reported in the physics literature 
\cite{Bucher1986, Wallingford}. 
As pointed out by Bejan \cite{Bejan1994}, 
these instances mirrored the almost parallel developments
in the techniques of ``entropy generation minimization''
amongst the engineering community and that of 
``finite-time thermodynamics'' within the physics community.
It is interesting to note that these earlier 
diagrams show changes in entropy by
angles, whereas  
the present diagram depicts such changes by 
line segments, while making use of the thermodynamic 
equilibrium curve of the finite-system involved.


\begin{thebibliography}{10}
%
\bibitem{Callenbook} H. B. Callen,
{\it Thermodynamics and an introduction 
to thermostatistics}, 2nd ed. John Wiley \& Sons Inc. (1985). 
%
 \bibitem{Bucher1993}
 Manfred Bucher; Diagram of the second law of thermodynamics. Am. J. Phys. 1 May 1993; 61 (5): 462–466. https://doi.org/10.1119/1.17242
%
\bibitem{Vallejo24a}
Andrés Vallejo; A diagrammatic representation of entropy production. Am. J. Phys. 1 March 2024; 92 (3): 234–235. https://doi.org/10.1119/5.0167570
%
\bibitem{Brodinskii1973}
V. M. Brodyansky. {\it The exergy method of thermodynamic analysis}, Moscow: Energy; 1973 (in Russian).
 
\bibitem{Bejan1977}
A. Bejan; Graphic techniques for teaching engineering thermodynamics, 
Mechanical Engineering News, pp. 26-28, May 1977.
%
\bibitem{Bucher1986}
Manfred Bucher; New diagram for heat flows and work in a Carnot cycle. Am. J. Phys. 1 September 1986; 54 (9): 850–851. https://doi.org/10.1119/1.14431
 \bibitem{Wallingford}
 J. Wallingford; Inefficiency and irreversibility in the Bucher diagram. Am. J. Phys. 1 April 1989; 57 (4): 379–381. https://doi.org/10.1119/1.16030
 %

\bibitem{Bejan1994}
A. Bejan, "Engineering Advances on Finite-Time Thermodynamics," Am. J.
Phys. 1 January 1994; 62 (1): 11-12.



\end{thebibliography}
\end{document}